\documentclass[prl,aps,preprintnumbers,showpacs,superscriptaddress]{revtex4}
\usepackage{amsfonts,amssymb,amsmath}            
\usepackage{graphics,graphicx,epsfig}            
\usepackage{amsthm}                              

\def\openone{\leavevmode\hbox{\small1\normalsize\kern-.33em1}}

\newcommand{\tr}[1]{\mbox{Tr} \, #1 }
\newcommand{\be}{\begin{equation}}
\newcommand{\ee}{\end{equation}}
\newcommand{\bea}{\begin{eqnarray}}
\newcommand{\eea}{\end{eqnarray}}

\newcommand \s {\widetilde{s}}

\def\reff#1{(\ref{#1})}
\def\bbbc{{\mathchoice {\setbox0=\hbox{$\displaystyle\rm C$}\hbox{\hbox
to0pt{\kern0.4\wd0\vrule height0.9\ht0\hss}\box0}}
{\setbox0=\hbox{$\textstyle\rm C$}\hbox{\hbox
to0pt{\kern0.4\wd0\vrule height0.9\ht0\hss}\box0}}
{\setbox0=\hbox{$\scriptstyle\rm C$}\hbox{\hbox
to0pt{\kern0.4\wd0\vrule height0.9\ht0\hss}\box0}}
{\setbox0=\hbox{$\scriptscriptstyle\rm C$}\hbox{\hbox
to0pt{\kern0.4\wd0\vrule height0.9\ht0\hss}\box0}}}}
\newtheorem{lemma}{Lemma}
\newtheorem{theorem}{Theorem}

\def\bbbc{{\rm I\!C}}

\def \iden {\bf{I}}

\begin{document}

\title{A quantum channel with additive minimum output entropy}


\author{Nilanjana \surname{Datta}}
\email[]{n.datta@statslab.cam.ac.uk}
\affiliation{Statistical Laboratory,
             Centre for Mathematical Science,
             University of Cambridge,
             Wilberforce Road,
             Cambridge CB3 0WB, UK }

\author{Alexander S. \surname{Holevo}}
\email[]{holevo@mi.ras.ru}
\affiliation{Steklov Mathematical Institute,
             Gubkina 8,
             119991 Moscow,
             Russia}

\author{Yuri \surname{Suhov}}
\email[]{yms@statslab.cam.ac.uk}
\affiliation{Statistical Laboratory,
             Centre for Mathematical Science,
             University of Cambridge,
             Wilberforce Road,
             Cambridge CB3 0WB, UK }


\begin{abstract}
We give a direct proof of the additivity of the minimum output entropy of a
particular quantum channel which breaks the multiplicativity conjecture.
This yields additivity of the classical capacity of this channel, a result
obtained by a different method in \cite{my}. Our proof relies heavily upon
certain concavity properties of the output entropy which are of independent
interest.
\end{abstract}


\maketitle

\bigskip

\section{Introduction}

\label{intro} A number of important issues of quantum information theory
would be greatly clarified if several resources and parameters were proved
to be additive. However, the proof of additivity of such resources
as the minimum output entropy of a quantum memoryless channel and its
classical capacity remains in general an open problem, see e.g.\,\cite{hol}.
Recently Shor \cite{Shor} provided a new insight into how several
additivity--type properties are related to each other. He proved that:
{(i)} additivity of the minimum output entropy of a quantum
channel, {(ii)} additivity of the classical capacity of a quantum channel,
{(iii)} additivity of the
entanglement of formation, and {(iv)} strong superadditivity of the
entanglement of formation are equivalent in the sense that if one of them
holds for
all channels then the others also hold for all channels.

In this paper we study the additivity of the minimum output entropy for a
channel which is particularly interesting because it breaks a closely
related multiplicativity property \cite{HW}. For this channel the
additivity of the classical capacity and of the minimum output entropy are
equivalent, which allows us to derive
an alternative proof of the result in \cite{my},
where additivity of its capacity was established. The problem
of additivity of the minimum output entropy is interesting and important in
its own right (it is straightforward, addresses a fundamental geometric
feature of a channel and may provide insight into more complicated channel
properties).

In this paper, the key observation that ensures the additivity is that the
output entropy of the product channel exhibits specific concavity properties
as a function of the Schmidt coefficients of the input pure state. It is our
hope that a similar mechanism might be responsible for the additivity of the
minimum output entropy in other interesting cases.

\section{The additivity conjecture}

\label{additivity}

A channel $\Phi $ in the finite dimensional Hilbert space ${\mathcal{H}}\simeq{\mathbf{C}}^{d}$
is a linear trace-preserving completely positive map of
the $\ast -$algebra  of complex $d\times d-$matrices.
A state is a density matrix $\rho ,$ that is Hermitian matrix such that $
\rho \geq 0,$ $\mathrm{Tr}\rho =1.$ The minimum output entropy of the
channel is defined as
\begin{equation}
h(\Phi ):=\min_{\rho }\,S(\Phi (\rho )),  \label{minent1}
\end{equation}
where the minimization is over all possible input states of the channel.
Here $S(\sigma )=-$ $\sigma \log \sigma $ is the von Neumann entropy of the
channel output matrix $\sigma =\Phi (\rho )$. The additivity problem for the
minimum output entropy is to prove that
\begin{equation}
h(\Phi _{1}\otimes \Phi _{2})=h(\Phi _{1})+h(\Phi _{2}),  \label{am}
\end{equation}
where $\Phi _{1},\Phi _{2}$ are two channels in ${\mathcal{H}_1},{\mathcal{H}_2}$ respectively,
$\otimes $ denotes tensor product.

A channel $\Phi $ is \textit{covariant}, if there are unitary representations $U_g, V_g$ of a group $G$ such that
\begin{equation}
\Phi (U_g \rho U_g^*)=V_g\Phi (\rho )V_g^*;\quad g\in G.
\end{equation}
If both representations are irreducible, then we call the channel irreducibly covariant. In this case there is a simple formula
\begin{equation}
\bar{C}(\Phi )=\log d-h(\Phi ),\label{hc}
\end{equation}
relating the Holevo capacity $\bar{C}(\Phi )$ of the channel with $h(\Phi )$ \cite{hol1}.
Since the tensor product of irreducibly covariant channels (with respect to possibly different groups $G_1, G_2$) is again irreducibly covariant (with respect to the group $G_1\times G_2$)),
it follows that if (\ref{am}) holds for two such channels, then
\begin{equation}
\bar{C}(\Phi _{1}\otimes \Phi _{2})=\bar{C}(\Phi _{1})+\bar{C}(\Phi _{2}).
\label{ahc}
\end{equation}
Notice that this does not follow from the result of \cite{Shor} which asserts that if (\ref{am})
holds for \textit{all} channels, then (\ref{ahc}) also holds for all channels. 
In the latter case also $\bar{C}(\Phi _{1}\otimes \dots \otimes \Phi _{n})=\bar{C}(\Phi
_{1})+\dots +\bar{C}(\Phi _{n}),$ which implies that $\bar{C}(\Phi )$ is
equal to the classical capacity of the channel $\Phi $ (see \cite{hol} for
more detail).

The concavity of the von Neumann entropy implies that the minimization in
\reff{minent1} can be restricted to pure input states, since the latter
correspond to the extreme points of the convex set of input states. Hence,
we can equivalently write the minimum output entropies in the form
\begin{eqnarray}
h(\Phi ) &=&\min_{{{|\psi \rangle \in \mathcal{H}}}\atop{{||\psi ||=1}}
}\,S(\Phi (|\psi \rangle \langle \psi |));\quad   \label{entrop1} \\
h(\Phi _{1}\otimes \Phi _{2}) &=&
\min_{{{|\psi_{12} \rangle \in {\cal{H}}_1 \otimes {\cal{H}}_2  }}\atop{{
||\psi_{12} ||=1}}}\,
S((\Phi
_{1}\otimes \Phi _{2})(|\psi _{12}\rangle \langle \psi _{12}|)).
\label{entrop2}
\end{eqnarray}

Here $|\psi _{12}\rangle \langle \psi _{12}|$ is a pure state of a
bipartite system with the Hilbert space ${\mathcal{H}}_{1}\otimes {\mathcal{H
}}_{2}$,
where ${\mathcal{H}}_{i}\simeq {\mathbf{C}}^{d_i}$ for $i=1,2$. 
In order to prove (\ref{am}), it is sufficient to show that the minimum in
(\ref{entrop2}) is attained on unentangled vectors $|\psi _{12}\rangle$.
 Consider the
Schmidt decomposition
\begin{equation}
|\ \psi _{12}\rangle =\sum_{\alpha =1}^{d}\sqrt{\lambda _{\alpha }}|\alpha ;1
\rangle |\alpha ; 2 \rangle,  \label{schmidt}
\end{equation}
where $d=\min\{d_1,d_2\}$, $\left\{ |\alpha ; j \rangle \right\} $ is an orthonormal basis in ${\
\mathcal{H}}_{j};j=1,2$, and ${\underline{\lambda }}=(\lambda _{1},\ldots
,\lambda _{d})$ is the vector of the Schmidt coefficients. The state $|
\psi _{12}\rangle \langle \psi _{12}|$ can then be expressed as
\begin{equation}
|\psi _{12}\rangle \langle  \psi _{12}|=\sum_{\alpha ,\beta =1}^{d}\sqrt{
\lambda _{\alpha }\lambda _{\beta }}|\alpha ;1\rangle \langle \beta ;1
|\otimes |\alpha ;2 \rangle \langle \beta  ;2|.
\end{equation}
The Schmidt coefficients form a probability distribution:
\begin{equation}
\lambda _{\alpha }\geq 0\quad ;\quad \sum_{\alpha =1}^{d}\lambda _{\alpha
}=1,  \label{nine}
\end{equation}
thus the vector ${\underline{\lambda }}$ varies in the $(d-1)-$dimensional
simplex $\Sigma _{d}$, defined by these constraints.
Extreme points (vertices) of $\Sigma _{d}$ correspond precisely to
unentangled vectors $|\psi _{12}\rangle =|\psi _{1}\rangle \otimes |\psi
_{2}\rangle \in {\mathcal{H}}_{1}\otimes {\mathcal{H}}_{2}$. The proof of
(\ref{am}) becomes straightforward {if} we can prove that for every choice
of the bases, the function
\begin{equation}
\underline{\lambda }\in \Sigma _{d}\mapsto S(M({\ \underline{\lambda }}))
\label{map}
\end{equation}
attains its minimum at the vertices of $\Sigma _{d}.$ Here $S(M({\
\underline{\lambda }}))$ is the von Neumann entropy of the channel matrix
\begin{equation}
M({\underline{\lambda }}):=\left( \Phi _{1}\otimes \Phi _{2}\right) \left(
|\psi _{12}\rangle \langle \psi _{12}|\right)=\sum_{\alpha ,\beta =1}^{d}\sqrt{\lambda _{\alpha
}\lambda _{\beta }}\Phi _{1}(|\alpha ;1\rangle \langle \beta  ;1|)\otimes
\Phi _{2}(|\alpha ;2\rangle \langle \beta  ;2|).  \label{matrix}
\end{equation}

Two special properties of a function can guarantee this: one is concavity,
and another is Shur concavity (see the Appendix). Both of them appear useful in
consideration of the particular channel we pass to.

\section{The channel}

\label{channel1}

The channel considered in this paper was introduced in \cite{HW}. It
is defined by its action on $d\times d$ matrices $\mu $ as follows:
\begin{equation}
\Phi (\mu )=\frac{1}{d-1}\bigl(\iden\,\mathrm{tr}(\mu )-\mu ^{T}\bigr)
\label{channel}
\end{equation}
where $\mu ^{T}$ denotes the transpose of the matrix $\mu $, and $\iden$ is
the unit matrix in ${\mathcal{H}} \simeq {\mathbf{C}}^{d}$. It is easy to see that the map $\Phi $
is linear and trace-preserving. For the proof of complete positivity see  \cite{HW}.
Moreover, $\Phi $ is irreducibly covariant since for any arbitrary unitary transformation $U$
\begin{equation}
\Phi (U\mu U^{\ast })=\bar{U}\Phi (\mu )\bar{U}^{\ast },  \label{contra}
\end{equation}
hence the relation (\ref{hc}) holds for this channel.

Our aim will be to prove the additivity relation
\begin{equation}
h(\Phi \otimes \Phi )=2h(\Phi ),  \label{amoe}
\end{equation}
for the channel (\ref{channel}). For $d=2$, (\ref{channel}) is a unital
qubit channel, for which property (\ref{amoe}) follows from \cite{king}. For
$d\geq 3$, (\ref{amoe}) can be deduced from additivity of the Holevo capacity
(\ref{ahc}), established in \cite{my}, by a different method. Here we provide a
direct proof based on the idea described at the end of the previous section.

For a pure state $\rho =|\psi \rangle \langle \psi |$ the channel output is
given by
\begin{equation}
\Phi (|\psi \rangle \langle \psi |)=\frac{1}{d-1}\Bigl({\iden}-|
{\overline{\psi }}\rangle \langle {\ \overline{\psi }}|\Bigr), \label{phi}
\end{equation}
where the entries of vector $|{\overline{\psi }}\rangle $ are complex
conjugates of the corresponding entries of vector $|{\psi }\rangle $. The
matrix $\Phi (|\psi \rangle \langle \psi |)$ has a non-degenerate eigenvalue
equal to $0$ and an eigenvalue $1/(d-1)$ which is $(d-1)$--fold degenerate.
The von Neumann entropy $S(\Phi (|\psi \rangle \langle \psi |))$ is obviously
the same for all pure
states, and so
\begin{equation}
h(\Phi )=\log (d-1).  \label{minent}
\end{equation}
As argued in the previous section, in order to prove (\ref{amoe}),
it is sufficient to show that the minimum in
$$
h(\Phi \otimes \Phi )=\min_{{|\psi _{12}\rangle \in {\mathbf{C}}
^{d}\otimes {\mathbf{C}}^{d}}\atop{{||\psi _{12}||=1}}}\,S((\Phi \otimes \Phi
)(|\psi _{12}\rangle \langle \psi _{12}|))  \label{entro2}
$$
is attained on unentangled vectors $|\psi _{12}\rangle$.
Consider the Schmidt decomposition \reff{schmidt} of $|\psi _{12}\rangle$.
Owing to the property \reff{contra}, we can choose for $\left\{ |\alpha ; j\rangle, \right\}; j=1,2,$
the canonical basis in ${\mathbf{C}}^{d}$. As it was shown in the previous section,
it suffices
to check that $S(M({\underline{\lambda}}))$ attains its minimum at the
vertices of $\Sigma _{d}.$ Here $M({\underline{\lambda }})$ is the
matrix defined in \reff{matrix} for the channel under consideration:
$$
M({\underline{\lambda }})=\sum_{\alpha ,\beta =1}^{d}\sqrt{\lambda _{\alpha
}\lambda _{\beta }}\Phi (|\alpha \rangle \langle \beta |)\otimes \Phi
(|\alpha \rangle \langle \beta |),
$$
where by \reff{channel}
$$
\Phi \left(|\alpha \rangle \langle \beta |\right)
= \frac{1}{d-1} \left( \delta_{\alpha \beta} \iden -
|\beta \rangle \langle \alpha |\right),
$$
owing to the fact that $|\alpha \rangle$ and $|\beta \rangle $ are
real.

Using (\ref{schmidt}) and the completeness relations:
\[
\,{\iden}=\sum_{\alpha =1}^{d}|\alpha \rangle \langle \alpha |,\quad {\iden}
\otimes {\iden}=\sum_{\alpha ,\beta =1}^{d}|\alpha \beta \rangle \langle
\alpha \beta |,
\]
we obtain
\begin{equation}
M({\underline{\lambda }})=\frac{1}{(d-1)^{2}}\left[ \sum_{\alpha ,\beta
=1}^{d}|\alpha \beta \rangle \langle \alpha \beta |(1-\lambda _{\alpha
}-\lambda _{\beta })+\sum_{\alpha ,\beta =1}^{d}\sqrt{\lambda _{\alpha
}\lambda _{\beta }}|\alpha \alpha \rangle \langle \beta \beta |\right] .
\label{opschmidt}
\end{equation}

In order to find the eigenvalues of $M({\underline{\lambda }})$,
it is instructive to first study the secular equation
of a more general $n\times n$ matrix:
$$
A=\sum_{j=1}^{n}\mu _{j}|j\rangle \langle j|+\sum_{j,k=1}^{n}\sqrt{\eta
_{j}\eta _{k}}|j\rangle \langle k|.  \label{opa}
$$
Matrix $A$ gives $(\Phi \otimes \Phi )(|\psi _{12}\rangle \langle \psi
_{12}|)$ for a particular choice of the parameters $\mu _{j}$ and $\eta _{j}$
[see eq.(\ref{parameters}) below]. It has the form:
\begin{equation}
\left(
\begin{array}{ccccc}
\mu _{1}+\eta _{1} & \sqrt{\eta _{1}\eta _{2}} & \cdots & \cdots & \sqrt{
\eta _{1}\eta _{n}} \\
\sqrt{\eta _{2}\eta _{1}} & \mu _{2}+\eta _{2} & \sqrt{\eta _{2}\eta _{3}} &
\cdots & \sqrt{\eta _{2}\eta _{n}} \\
\vdots & \vdots & \vdots & \vdots & \vdots \\
\sqrt{\eta _{n}\eta _{1}} & \cdots & \cdots & \cdots & \mu _{n}+\eta _{n} \\
&  &  &  &
\end{array}
\right) .  \label{matrixa}
\end{equation}
The secular equation ${\hbox{det}}(A-\gamma {\iden})=0$
can be written as
\be
F(\gamma )=0,
\label{sec}
\ee
where
$$
F(\gamma )=\prod_{j}(\mu _{j}-\gamma )\Bigl[1+\frac{\eta _{1}}{\mu
_{1}-\gamma }+\ldots +\frac{\eta _{n}}{\mu _{n}-\gamma }\Bigr].  \label{dn}
$$
Solving eq.(\ref{sec}) would be in general non-trivial. However,
representing the matrix $\bigl[(d-1)^{2}M({\underline{\lambda }})\bigr]$
in the form (\ref{matrixa}) results in a convenient expression for
$F(\gamma )$. This allows us to identify many of the eigenvalues of
$(d-1)^{2}$ $M({\underline{
\lambda }})$. More precisely, we identify $j$ with a pair $(\alpha ,\beta )$
and obtain
\begin{equation}
\mu _{j}\equiv \mu _{\alpha \beta }=1-\lambda _{\alpha }-\lambda _{\beta
}\quad ;\quad \eta _{j}\equiv \eta _{\alpha \beta }=\lambda _{\alpha }\delta
_{\alpha \beta },\quad \alpha ,\beta =1,\ldots ,d.  \label{parameters}
\end{equation}
Therefore,
\begin{eqnarray}
F(\gamma ) &=&\prod_{\alpha ,\beta =1}^{d}(1-\lambda _{\alpha }-\lambda
_{\beta }-\gamma )\left[1+\sum_{\alpha ^{\prime },\beta ^{\prime }=1}^{d}
\frac{\lambda _{\alpha ^{\prime }}\delta _{{\alpha ^{\prime }}{\beta
^{\prime }}}}{(1-\lambda _{\alpha ^{\prime }}-\lambda _{\beta ^{\prime
}}-\gamma )}\right]  \nonumber  \label{long} \\
&=&\prod_{{{\alpha ,\beta =1}}{{\alpha \neq \beta }}}^{d}(1-\lambda
_{\alpha }-\lambda _{\beta }-\gamma )\left[\prod_{\alpha ^{\prime
}=1}^{d}(1-2\lambda _{\alpha ^{\prime }}-\gamma )\left\{1+\sum_{\alpha
^{^{\prime \prime }}=1}^{d}\frac{\lambda _{\alpha ^{^{\prime \prime }}}}{
(1-2\lambda _{\alpha ^{^{\prime \prime }}}-\gamma )}\right\}\right].  \nonumber
\\
&&
\end{eqnarray}

Eq.(\ref{sec}) yields the following equations:
\begin{equation}
(1-\lambda _{\alpha }-\lambda _{\beta }-\gamma )=0,\quad \alpha \neq \beta
,\quad \alpha ,\beta = 1,2,\ldots ,d,  \label{one1}
\end{equation}
where $\lambda _{\alpha },\lambda _{\beta }$ denote the Schmidt coefficients
(\ref{schmidt}). Equation (\ref{one1}) implies that there are $d(d-1)$
eigenvalues of the form
\begin{equation}
\gamma =1-\lambda _{\alpha }-\lambda _{\beta }, \quad \alpha \ne \beta \quad
\alpha, \beta =1, \ldots, d.  \label{eigen1}
\end{equation}

The roots of the equation
\begin{equation}
\prod_{\alpha =1}^{d}(1-2\lambda _{\alpha }-\gamma )\left\{1+\sum_{\alpha'
=1}^{d}\frac{\lambda _{\alpha' }}{(1-2\lambda _{\alpha' }-\gamma )}\right\}=0
\label{two}
\end{equation}
give the remaining $d$ eigenvalues of the matrix
$\left[(d-1)^{2} M({\underline{\lambda }})\right]$.

For the case $d=3$ the roots of (\ref{two}) can be explicitly evaluated.
This is done in the next section. The case of arbitrary $d>3$ is discussed
in sections that follow. Note that the sum of all eigenvalues of $
\left[(d-1)^{2}M({\underline{\lambda }})\right]$ equals
$$
\mathrm{tr}\left[(d-1)^{2}M({\underline{\lambda }})\right]=
(d-1)^{2}\,\tr M({
\underline{\lambda }})=(d-1)^{2},
$$
since $M({\underline{\lambda }})$ is a density matrix acting in ${\mathbf{C}}
^{d^{2}}$.

\section{Eigenvalues for $d=3$}

\label{3d}

For $d=3$, there are $d(d-1)=6$ eigenvalues of the matrix $\left[(d-1)^{2} M({
\underline{\lambda }})\right]=4 M({\underline{\lambda }})$, which are given by
\reff{eigen1}. The sum of these eigenvalues is:
$$
\sum_{\alpha ,\beta = 1 \atop{\alpha \neq \beta }}^3 (1-\lambda _{\alpha
}-\lambda _{\beta })=2\bigl[3-2(\lambda _{1}+\lambda _{2}+\lambda _{3})\bigr]
=2  \label{first}
$$
since
\begin{equation}
\lambda _{1}+\lambda _{2}+\lambda _{3}=1.  \label{sum}
\end{equation}
The remaining three eigenvalues of $4M({\underline{\lambda }})$ are given
by the roots of the equation
\begin{equation}
\prod_{\alpha =1}^{3}(1-2\lambda _{\alpha }-\gamma )\left\{1+\sum_{\alpha'
=1}^{3}\frac{\lambda _{\alpha' }}{(1-2\lambda _{\alpha' }-\gamma )}\right\}=0.
\label{two2}
\end{equation}
Since the sum of all the eigenvalues is equal to $(d-1)^{2}\equiv 4$, these
remaining three eigenvalues sum up to $4-2=2$. Using (\ref{sum}), we can
cast (\ref{two2}) as:
\begin{equation}
\gamma ^{3}+a_{2}\gamma ^{2}+a_{1}\gamma +a_{0}=0  \label{cubic}
\end{equation}
where
\begin{equation}
a_{0}=-4\lambda _{1}\,\lambda _{2}\,\lambda
_{3}\,\,;\,\,a_{1}=1\,\,;\,\,a_{2}=-2.\,\,\,\,  \label{parameters2}
\end{equation}
The three roots of (\ref{cubic}) are given by
\begin{eqnarray}
{\widetilde{\gamma }}_{1}:= &&-\frac{a_{2}}{3}+(T_{1}+T_{2}),  \nonumber \\
{\widetilde{\gamma }}_{2}:= &&-\frac{a_{2}}{3}-\frac{1}{2}\,(T_{1}+T_{2})+
\frac{1}{2}\,i\sqrt{3}\,(T_{1}-T_{2}),  \nonumber \\
{\widetilde{\gamma }}_{3}:= &&-\frac{a_{2}}{3}-\frac{1}{2}\,(T_{1}+T_{2})-
\frac{1}{2}\,i\sqrt{3}\,(T_{1}-T_{2}).  \label{roots}
\end{eqnarray}
Here
$$
T_{1}:=\Bigl[R+\sqrt{D}\Bigr]^{1/3}\quad {\hbox{and}}\quad T_{2}:=\Bigl[R-
\sqrt{D}\Bigr]^{1/3},  \label{st}
$$
and
\begin{equation}
R=\frac{1}{54}\,(9a_{1}a_{2}-27a_{0}-2a_{2}^{3}), \quad
D=Q^{3}+R^{2}, \quad  Q:=\frac{1}{9}(3a_{1}-a_{2}^{2}).
\label{arr}
\end{equation}

Thus, the matrix $M({\underline{\lambda }})$
has six eigenvalues of the form $
(1/4)(1-\lambda _{\alpha }-\lambda _{\beta })$, where $\alpha ,\beta =1,2,3$
and $\alpha \neq \beta $, and three eigenvalues $\gamma _{1}$, $\gamma _{2}$
and $\gamma _{3}$, with $\gamma _{i}:={\widetilde{\gamma }}_{i}/4$.
The output entropy $S\bigl(M({\underline{\lambda }})\bigr)$ can be expressed
as the sum:
$$
S(M({\underline{\lambda }}))=S_{1}({\underline{\lambda }})+S_{2}({\underline{
\lambda }}).  \label{tot_entropy}
$$
Here
\begin{eqnarray}
S_{1}({\underline{\lambda }}) &=&-\sum_{\alpha ,\beta =1 \atop{\alpha
\neq \beta}}^{3}\frac{1}{4}(1-\lambda _{\alpha }-\lambda _{\beta })\log \left[
\frac{1}{4}(1-\lambda _{\alpha }-\lambda _{\beta })\right]  \nonumber \\
&=&-\frac{1}{2}\sum_{\alpha =1}^{3}\lambda _{\alpha }\log \frac{\lambda
_{\alpha }}{4}  = \frac{1}{2}H({\underline{\lambda }})+1  \label{ent1}
\end{eqnarray}
where $H({\underline{\lambda }})=-\sum_{\alpha =1}^{d}\lambda _{\alpha }\log
{\lambda _{\alpha }}$ denotes the Shannon entropy of ${\underline{
\lambda }}$, and
$$
S_{2} ({\underline{\lambda }})=
-\sum_{i=1}^{3}\gamma _{i}\log \gamma _{i}.  \label{ent2}
$$
Since $H({\underline{\lambda }})$ is a concave function of ${\underline{
\lambda }}=(\lambda _{1},\lambda _{2},\lambda _{3})$, so is $S_{1}({
\underline{\lambda }})$. Hence $S_{1}({\underline{\lambda }})$ attains its
minimum at the vertices of $\Sigma _{3}$.

Let us now evaluate the summand $S_{2}({\underline{\lambda }})$.
Substituting the values of $a_{0}$, $a_{1}$ and $a_{2}$ from (\ref
{parameters2}) into (\ref{arr}), we get
$$
R=-\frac{1}{27}+2t,\quad Q=-\frac{1}{9}, \quad D=-4t(\frac{1}{27}-t)\leq 0.
$$
Here $t=\lambda _{1}\lambda _{2}\lambda _{3},0\leq t\leq 1/27$. Hence, we
can write
$$
R+\sqrt{D}=R+i\sqrt{|D|}=re^{i\theta },
$$
where $r=\sqrt{R^{2}+|D|}=1/27$ and $\theta =\arctan (\sqrt{|D|}/R),0\leq
\theta \leq \pi ,$ so that
\[
\tan \theta =\frac{\sqrt{t(1/27-t)}}{t-1/54}.
\]
Considering the sign of this expression we find that $t=0$ corresponds to $
\theta =\pi ,$ while $t=1/27$ to $\theta =0$. In terms of $\theta $ the
eigenvalues $\gamma _{k}$, $k=1,2,3$  can now be expressed as:
$$
\gamma _{k}=\frac{1}{6}\left[ 1+\cos \left( \frac{\theta }{3}-\frac{2\pi
(k-1)}{3}\right) \right] =\frac{1}{3}\cos ^{2}\left( \frac{\theta }{6}-\frac{
2\pi (k-1)}{6}\right) .
$$
Hence,
\begin{equation}
S_{2}({\underline{\lambda }})=-\sum_{k=1}^{3}\frac{1}{3}\cos ^{2}\left(
\frac{\theta }{6}-\frac{2\pi (k-1)}{6}\right) \log \left[\frac{1}{3}\cos
^{2}\left( \frac{\theta }{6}-\frac{2\pi (k-1)}{6}\right)\right].
\label{long2}
\end{equation}

An argument similar to Lemma 3 of \cite{barnett} shows that the RHS of (\ref
{long2}) has a global minimum, equal to $1$, at $\theta =\pi $ corresponding
to $t=\lambda _{1}\lambda _{2}\lambda _{3}=0$. Hence,
$S_{2}({\underline{\lambda }})$ attains its
minimal value $1$ at every point of the boundary $\partial \Sigma _{3},$ in
particular at its vertices: $\lambda _{i}=1,\lambda _{j}=0$ for $j\neq i$, $
i=1,2,3.$ The summand $S_{1}({\underline{\lambda }})$, given by (\ref{ent1}
), also attains its minimum, equal to $1$, at the vertices. Therefore the
sum $S(M({\underline{\lambda }}))$ attains its minimum, equal to $2$, at the
vertices of $\Sigma _{3}.$ Hence, $h(\Phi \otimes \Phi )=2$, and the
additivity (\ref{amoe}) holds, as $h(\Phi )=1$ by (\ref{minent}).

We conjecture that the entropy $S(M({\underline{\lambda }}))$ as a function
of ${\underline{\lambda }}$ is concave. This is supported by a $3D$-plot of $
S(M({\underline{\lambda }}))$ as a function of two independent Schmidt
coefficients $\lambda _{1}$ and $\lambda _{2}$; here $\lambda _{i}\geq 0$
for $i=1,2$ and $\lambda _{1}+\lambda _{2}\leq 1$. See Figure 1 below.
\begin{figure}[h]
\centerline{\includegraphics{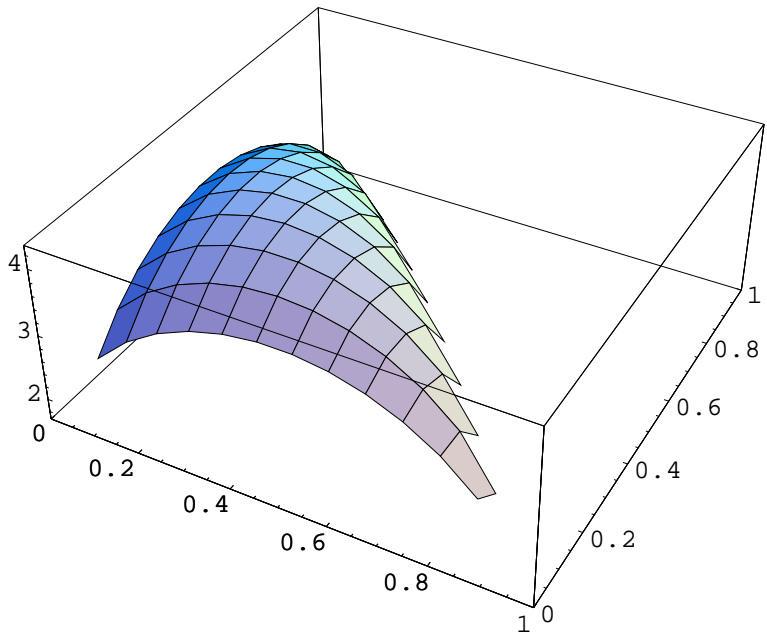}}
\centerline{\parbox{11cm}{\caption{ The entropy $S(M({\underline{\lambda }}))$
as a function of two independent Schmidt\\ coefficients $\lambda_1$ and
$\lambda_2$.}}}
\label{fig1}
\end{figure}

However $S_{2}({\underline{\lambda }})$ is not concave as can be seen
e.g. by taking $\lambda _{2}=\lambda _{1},$
$\lambda _{3}=1-$ 2$\lambda _{1}.$ See Figure 2.
\begin{figure}[h]
\centerline{\includegraphics{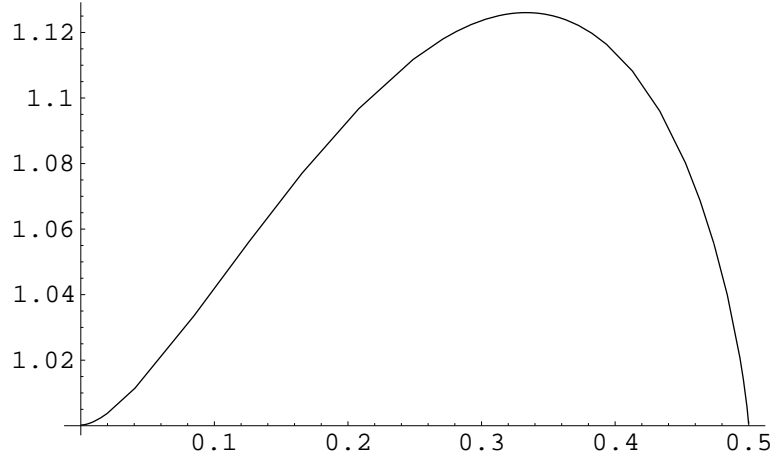}}
\centerline{\parbox{11cm}{\caption{$S_{2}({\underline{\lambda }})$
as a function of $\lambda=\lambda_1 = \lambda_2$.}}}
\label{fig2}
\end{figure}

\section{Minimum output entropy in $d>3$ dimensions}
\label{arbitraryd}
In a previous section we found that the matrix $
\left[(d-1)^{2} M({\underline{\lambda }})\right]$, where
$M({\underline{\lambda }})$ is the output density matrix of the
channel $\Phi \otimes \Phi$ and is given by \reff{opschmidt},
has $d(d-1)$ eigenvalues of the form
\begin{equation}
\left(1-\lambda _{\alpha }-\lambda _{\beta }\right),\quad {\hbox{with}} \quad
\alpha \neq \beta ,\,\,\alpha
,\beta =1,2,\ldots ,d,  \label{eigen11}
\end{equation}
and the remaining $d$ eigenvalues are given by the roots $\gamma _{1},\ldots
,\gamma _{d}$ of (\ref{two}). Hence, the matrix $M({\underline{\lambda }})$
has $d(d-1)$ eigenvalues of the form
$$
e_{\alpha \beta }:=\frac{1}{(d-1)^{2}}\bigl(1-\lambda _{\alpha }-\lambda
_{\beta }\bigr),\quad \alpha \neq \beta ,\,\,\alpha ,\beta =1,2,\ldots ,d,
$$
and $d$ eigenvalues of the form
\[
g_{i}:=\frac{\gamma _{i}}{(d-1)^{2}},\quad i=1,2,\ldots ,d.
\]
Note that the $\gamma _{i}$'s and $g_{i}$'s are functions of
${\underline{\lambda }} \in \Sigma _{d}$.
Accordingly, we write the von Neumann entropy of the output density matrix
as a sum
\begin{equation}
S(M({\underline{\lambda }}))=S_{1}({\underline{\lambda }})+S_{2}({\underline{
\lambda }})  \label{sum2}
\end{equation}
where
\begin{equation}
S_{1}({\underline{\lambda }}):=-\sum_{\alpha =1}^{d}\sum_{\beta =1\atop{
\beta \neq \alpha }}^{d}e_{\alpha \beta }\log e_{\alpha \beta },
\quad S_{2}({\underline{\lambda }}):=-\sum_{i=1}^{d}g_{i}\log g_{i}.
\label{s11}
\end{equation}
Note that
\begin{equation}
\sum_{\alpha =1}^{d}\sum_{\beta =1\atop{\beta \neq \alpha }}^{d}e_{\alpha \beta }=\sum_{\alpha =1}^{d}\sum_{\beta =1\atop{\beta
\neq \alpha }}^{d}\frac{1-\lambda _{\alpha }-\lambda _{\beta }}{(d-1)^{2}}=
\frac{d-2}{d-1}.  \label{sum33}
\end{equation}
Define the following variables:
\begin{eqnarray}
{\widetilde{e}}_{\alpha \beta }\,\left(={\widetilde{e}}_{\alpha \beta}(
{\underline{\lambda }})\right)\, &:=&\frac{d-1}{d-2}\,e_{\alpha \beta }=\frac{1
}{\left( d-1\right) \left( d-2\right) }\sum_{1\le \delta \le d\atop{\delta
\not=\alpha ,\beta
}}\lambda _{\delta },\quad \alpha \not=\beta, \quad \alpha, \beta = 1, \ldots,
d.  \label{aff} \\
\tilde{g}_{i}\,\left(= \tilde{g}_{i}({\underline{\lambda }})\right) \,&:=&\left( d-1\right) g_{i}=\frac{1}{\left( d-1\right) }\gamma
_{i}, \quad i=1,\ldots,d.
\label{tilg}
\end{eqnarray}
For $d\geq 3$ we have ${\widetilde{e}}_{\alpha \beta }\geq 0$, and from (\ref
{sum33}) it follows that
$$
\sum_{\alpha =1}^{d}\sum_{\beta =1\atop{\beta \neq \alpha }}^{d}{
\widetilde{e}}_{\alpha \beta } =\frac{d-1}{d-2}\sum_{\alpha =1}^{d}\sum_{
\beta =1\atop{\beta \neq \alpha }}^{d}e_{\alpha \beta }=1,
\quad
\sum_{i=1}^{d}\tilde{g}_{i} = \left( d-1\right) \left[ 1-\frac{d-2}{d-1}
\right] =1.
$$

Hence, ${\underline{{\widetilde{e}}}}:=\{{\widetilde{e}}_{\alpha \beta
}\,|\,\alpha \neq \beta ,\alpha ,\beta =1,2,\ldots ,d\}$ and ${\underline{{
\widetilde{g}}}}=\left\{ \tilde{g}_{i}|i=1,\dots ,d\right\}$ are
probability distributions. In terms of these variables
\begin{equation}
S_{1}({\underline{\lambda }})=\frac{d-2}{d-1}H({\underline{{\widetilde{e}}}}
)-\frac{d-2}{d-1}\log \left(\frac{d-2}{d-1}\right),  \label{s222}
\end{equation}
where $H({\underline{{\widetilde{e}}}})$ denotes the Shannon entropy of ${
\underline{{\widetilde{e}}}}$. In view of (\ref{s11})
\be
S_{2}({\underline{\lambda }})=\frac{1}{\left( d-1\right) }
H({\underline{{
\widetilde{g}}}})+\frac{1}{d-1}\log \left( d-1\right) .
\label{summ2}
\ee

From (\ref{s222}) it follows that $S_{1}({\underline{\lambda }})$
in (\ref{sum2}) is a concave function of the variables ${\widetilde{e}}
_{\alpha \beta }$. These variables are affine
functions of the Schmidt coefficients $\lambda_1, \ldots, \lambda_d$.
Hence, $S_{1}$ is
a concave function of $\ {\underline{\lambda }}$ and attains its minimum at
the vertices of $\Sigma _{d}$, defined by the constraints \reff{nine}.

Let us now analyze $S_{2}({\underline{\lambda }})$. We wish to
prove the following:
\begin{theorem}
The function $S_{2}$ is Schur-concave in $\underline{
\lambda }\in \Sigma _{d}$ i.e., $\underline{\lambda }\prec
\underline{\lambda }^{\prime }\, \implies S_{2}\left( {
\underline{\lambda }}\right) \geq S_{2}\left( {\underline{\lambda }}^{\prime
}\right)$, where $\prec $ {denotes the majorization order}  (see
the Appendix).
\end{theorem}

Since every $\underline{\lambda }\in \Sigma _{d}$ is majorized by the
vertices of $\Sigma _{d}$, this will imply that $S_{2}(\underline{\lambda })$ also attains its
minimum at the vertices. Thus $S(\underline{\lambda })=S_{1}(\underline{\lambda })+S_{2}(\underline{\lambda })$ is minimized at the vertices,
which correspond to unentangled states. As was observed, this
implies the additivity.

\section{Proof of the Theorem}

We will use the quite interesting observation made in \cite{graeme}, that the Shannon entropy $H({
\underline{x}})$ is a monotonically increasing function of the elementary
symmetric polynomials $s_{k}(x_{1},x_{2},\ldots ,x_{d}),$ $k=0,\ldots,
d$,
in the variables ${\underline{x}}=\left( x_{1},x_{2},\ldots ,x_{d}\right)$.
The latter are
defined by equations \reff{sym2a} of the Appendix.
Hence the Shannon entropy $H({\underline{{\widetilde{g}}}})$
in \reff{summ2} is a monotonically increasing function of the
symmetric polynomials
\begin{equation}
\s_{k}({\underline{\lambda}}):=s_{k}(\tilde{g}_{1},\tilde{g}_{2},\ldots ,
\tilde{g}_{d}) \equiv \frac{1}{(d-1)^k} s_{k}(\gamma _{1},\gamma _{2},\ldots
,\gamma _{d}),\quad k=0,\ldots,d,
\label{wides}
\end{equation}
Therefore, to prove the Theorem it is sufficient to prove
that the functions $\s_{k}({\underline{\lambda}})$ are
Schur concave in $\underline{\lambda }\in \Sigma _{d}$.
Here the variables $\tilde{g}_{i}$ are given by \reff{tilg},
and the variables $\gamma _{i}$ are the roots of eq. \reff{two}.
Define the variables
$$
\nu _{\alpha }:=1-2\lambda _{\alpha },\quad \alpha =1,2,\ldots ,d.
\label{nu}
$$
Note that $-1\leq \nu _{\alpha }\leq 1$, owing to the inequality $0\leq
\lambda _{\alpha }\leq 1$. Moreover,
$$
\sum_{\alpha =1}^{d}\nu _{\alpha }=d-2,  \label{sum3}
$$
since $\sum_{\alpha =1}^{d}\lambda _{\alpha }=1.$ In terms of the variables $
\nu _{\alpha }$, (\ref{two}) can be expressed as
\begin{equation}
\prod_{\alpha =1}^{d}(\nu _{\alpha }-\gamma )\left\{1+\frac{1}{2}
\sum_{\alpha' =1}^{d}\frac{1-\nu _{\alpha' }}{(\nu _{\alpha' }-\gamma )}
\right\}=0.  \label{twonu}
\end{equation}
Since the roots $\gamma_1, \ldots, \gamma_d$ of \reff{two} are identified,
trivially, as the zeroes of the product
$
(\gamma_1 - \gamma)(\gamma_2 - \gamma) \ldots (\gamma_d - \gamma),
$
equation \reff{twonu} can be expressed in terms of these roots as follows:
\begin{equation}
\sum_{k=0}^{d}\gamma ^{k}\,(-1)^{k}\,s_{d-k}(\gamma _{1},\gamma _{2},\ldots
,\gamma _{d}) =0. \label{sym1}
\end{equation}
In terms of the elementary symmetric polynomials $s_{l}$, of the
variables $\nu _{1},\nu _{2},\ldots ,\nu
_{d}$, (\ref{twonu}) can be rewritten as
\begin{equation}
\sum_{k=0}^{d}\gamma ^{k}\,(-1)^{k}\,s_{d-k}(\nu _{1},\nu _{2},\ldots ,\nu
_{d})+\sum_{k=0}^{d-1}\gamma ^{k}\,(-1)^{k}\,\sum_{l=1}^{d}s_{d-1-k}(\nu
_{1},\ldots ,{\not{\nu _{l}}}\ldots ,\nu _{d})\,\frac{(1-\nu _{l})}{2}=0,
\label{sym3}
\end{equation}
where the symbol ${\not{\nu _{l}}}$ means that the variable ${\nu _{l}}$ has
been omitted from the arguments of the corresponding polynomial.
Equating the LHS of (\ref{sym1}) with the LHS of (\ref{sym3}) yields,
for each $0\leq k\leq d-1$ :
\begin{equation}
s_{d-k}(\gamma _{1},\gamma _{2},\ldots ,\gamma _{d})=s_{d-k}(\nu _{1},\nu
_{2},\ldots ,\nu _{d})+\sum_{l=1}^{d}s_{d-1-k}(\nu _{1},\ldots ,{\not{\nu
_{l}}}\ldots ,\nu _{d})\,\frac{(1-\nu _{l})}{2}.  \label{sympol}
\end{equation}
Note that in (\ref{sympol}), values $s_{d-k}(\gamma _{1},\gamma
_{2},\ldots ,\gamma _{d})$ are expressed in terms of values of elementary
symmetric polynomials in the variables $\nu _{1},\nu _{2},\ldots ,\nu _{d}$
(which are themselves linear functions of the Schmidt coefficients $\lambda
_{1},\ldots ,\lambda _{d}$).

Our aim is to prove that $\s_{k}(\underline{\lambda })$ is Schur concave in
the Schmidt
coefficients $\lambda _{1},\ldots ,\lambda _{d}$. Eq.\reff{wides}
implies that this amounts to
proving Schur concavity of $s_{d-k}(\gamma _{1},\gamma _{2},\ldots ,
\gamma _{d})$ as a function of
$\lambda _{1},\ldots ,\lambda _{d}$, for all $0 \le k \le d$.
The functions
\begin{equation}
\Phi _{k}(\nu _{1},\ldots ,\nu _{d}):=s_{d-k}(\nu _{1},\ldots ,\nu
_{d})+\sum_{l=1}^{d}s_{d-1-k}(\nu _{1},\ldots ,{\not{\nu _{l}}}\ldots ,\nu
_{d})\,\frac{(1-\nu _{l})}{2}\equiv {\hbox{RHS of }}(\ref{sympol})
\label{phidef}
\end{equation}
are symmetric in the variables $\nu _{1},\nu _{2},\ldots ,\nu _{d}$, and
hence in the variables
$\lambda _{1},\ldots ,\lambda _{d}$. By eq.(\ref{iden3}) (see the Appendix) it
remains to prove
\be
(\lambda _{i}-\lambda
_{j})\bigl(\frac{\partial \Phi _{k}}{\partial \lambda _{i}}-\frac{\partial
\Phi _{k}}{\partial \lambda _{j}}\bigr) \equiv (\nu
_{i}-\nu _{j})\bigl(\frac{\partial \Phi _{k}}{\partial \nu _{i}
}-\frac{\partial \Phi _{k}}{\partial \nu _{j}}\bigr)\leq 0,\quad \forall
\,1\leq i,j\leq d.  \label{key2}
\ee
By (\ref{iden1}) we
have
\begin{eqnarray}
\frac{\partial }{\partial \nu _{i}}\Phi _{k}(\nu
_{1},\ldots ,\nu _{d}) &=& \frac{\partial }{\partial \nu _{i}}s_{d-k}(\nu
_{1},\ldots ,\nu _{d})+\frac{ \partial }{\partial \nu
_{i}}\sum_{l=1}^{d}s_{d-1-k}(\nu _{1},\ldots ,{\not {\nu _{l}}}\ldots ,\nu
_{d})\,\frac{(1-\nu _{l})}{2}  \nonumber \\ &=&s_{d-1-k}(\nu
_{1},..,{\not{\nu _{i}}},..,\nu _{d})+\sum_{{l=1 \atop{l\neq
i}}}^{d}s_{d-1-k}(\nu _{1},\ldots ,{\not{\nu _{i}}},..,{\not{\nu _{l}}}
\ldots ,\nu _{d})\,\frac{(1-\nu _{l})}{2}  \nonumber \\
&&\quad -\frac{1}{2}s_{d-1-k}(\nu _{1},\ldots ,{\not{\nu _{i}}}\ldots ,\nu
_{d}).  \label{eq1}
\end{eqnarray}
Therefore,
\begin{eqnarray}
\Bigl(\frac{\partial \Phi _{k}}{\partial \nu _{i}}-\frac{\partial \Phi _{k}}{
\partial \nu _{j}}\Bigr)(\nu _{1},\ldots ,\nu _{d}) &=&s_{d-1-k}(\nu _{1},..,
{\not{\nu _{i}}},..,\nu _{d})-s_{d-1-k}(\nu _{1},..,{\not{\nu _{j}}},..,\nu
_{d})  \nonumber \\
&+&\sum_{{l=1\atop{l\neq i}}}^{d}s_{d-1-k}(\nu _{1},..,{\not{\nu _{i}}}
,..,{\not{\nu _{l}}}..,\nu _{d})\,\frac{(1-\nu _{l})}{2}-\sum_{{l=1\atop{
l\neq j}}}^{d}s_{d-1-k}(\nu _{1},..,{\not{\nu _{j}}},..,{\not{\nu _{l}}}
\ldots ,\nu _{d})\,\frac{(1-\nu _{l})}{2}  \nonumber \\
&-&\frac{1}{2}\bigl[s_{d-1-k}(\nu _{1},\ldots ,{\not{\nu _{i}}}\ldots ,\nu
_{d})-s_{d-1-k}(\nu _{1},\ldots ,{\not{\nu _{j}}}\ldots ,\nu _{d})\bigr].
\end{eqnarray}
Using (\ref{iden2}) we get
\begin{eqnarray}
\Bigl(\frac{\partial \Phi _{k}}{\partial \nu _{i}}-\frac{\partial \Phi _{k}}{
\partial \nu _{j}}\Bigr)(\nu _{1},\ldots ,\nu _{d}) &=&\frac{1}{2}\bigl(\nu
_{j}-\nu _{i})s_{d-k-2}(\nu _{1},..,{\not{\nu _{i}}},..,{\not{\nu _{j}}}
\ldots ,\nu _{d})  \nonumber \\
&+&\frac{(\nu _{i}-\nu _{j})}{2}s_{d-k-2}(\nu _{1},..,{\not{\nu _{i}}},..,{
\not{\nu _{j}}}\ldots ,\nu _{d})  \nonumber \\
&+&\sum_{{l=1}\atop{{l\neq i,j}}}^{d}\frac{(1-\nu _{l})}{2}\bigl[
s_{d-k-2}(\nu _{1},..,{\not{\nu _{i}}},..,{\not{\nu _{l}}}\ldots ,\nu
_{d})-s_{d-k-2}(\nu _{1},..,{\not{\nu _{j}}},..,{\not{\nu _{l}}}\ldots ,\nu
_{d})\bigr]  \nonumber \\
&=&\sum_{{l=1}\atop{{l\neq i,j}}}^{d}\frac{(1-\nu _{l})}{2}\bigl[
s_{d-k-2}(\nu _{1},..,{\not{\nu _{i}}},..,{\not{\nu _{l}}}\ldots ,\nu
_{d})-s_{d-k-2}(\nu _{1},..,{\not{\nu _{j}}},..,{\not{\nu _{l}}}\ldots ,\nu
_{d})\bigr]  \nonumber \\
&=&\sum_{{l=1 \atop{l\neq i,j}}}^{d}\frac{(1-\nu _{l})}{2}(\nu _{j}-\nu
_{i})s_{d-k-3}(\nu _{1},..,{\not{\nu _{i}}},..,{\not{\nu _{j}}},..{\not{\nu
_{l}}}\ldots ,\nu _{d}).  \label{long22}
\end{eqnarray}
Substituting (\ref
{long22}) in (\ref{key2}) we obtain that Schur
concavity holds if and only if
\begin{equation}
\sum_{{l=1\atop{l \neq i,j}}}^{d}(1-\nu _{l})s_{d-k-3}(\nu _{1},..,{\not
{\nu _{i}}},..,{\not{\nu _{j}}},..{\not{\nu _{l}}}\ldots ,\nu _{d})\geq
0,\quad \forall \,\,1\leq i,j\leq d.  \label{main}
\end{equation}
The variables $\nu _{i}$ and $\nu _{j}$ do not appear in (\ref{main}). Owing
to symmetry, without loss of generality, we can choose $i=d-1$ and $j=d$.
Then omitting $\nu _{d-1}$ and $\nu _{d}$ results in replacing (\ref{main})
by
$$
\sum_{l=1}^{d-2}(1-\nu _{l})s_{d-k-3}(\nu _{1},..,{\not{\nu _{l}}}\ldots
,\nu _{d-2})\geq 0.
\label{ineq33}
$$
By setting $n=d-2$, we can express the condition for Schur concavity by the
following lemma.
\begin{lemma}
The functions $\Phi _{k}$, defined in \reff{phidef}, are Schur concave in
the Schmidt coefficients $\lambda_1, \ldots, \lambda_d$ if
\begin{equation}
\sum_{l=1}^{n}(1-\nu _{l})s_{n-k-1}(\nu _{1},..,{\not{\nu _{l}}}\ldots ,\nu
_{n})\geq 0,\quad  0 \le k \le d-3,   \label{main2}
\end{equation}
where the variables $\nu _{i} := 1 - 2 \lambda_i$, $1\le i \le n$, satisfy
\begin{equation}
-1\leq \nu _{i}\leq 1,  \quad \sum_{l=1}^{n}\nu _{l} \geq n-2.
\label{1nu1}
\end{equation}
\end{lemma}
\noindent
{\bf{Note}}: The constraints \reff{1nu1} follow from the
relations:\,
$\displaystyle{ \lambda_l \ge 0 \,\, \forall \, \, l}$, and  $\displaystyle{\sum_{l=1}^{n}\lambda _{l}=\sum_{l=1}^{d-2}\lambda _{l}\leq 1.}$
\bigskip

\noindent
{\bf{Proof of the Lemma}}

The constraints \reff{1nu1} imply that
\emph{at most one} of the
variables $\nu _{1},\ldots ,\nu _{n}$ can be negative. Note that $(1-\nu
_{l})$ is always nonnegative since $\nu _{l}\leq 1$. Thus if {all $\nu
_{1},\ldots ,\nu _{n}\geq 0,$ (\ref{main2}) obviously holds. }Hence, we need
to prove (\ref{main2}) only in the case in which one, and only one, of the
variables $\nu _{1},\ldots ,\nu _{n}$ is negative.

To establish the latter fact, we first prove the inequality
\begin{equation}
\sum_{l=1}^{n}\frac{1-\nu _{l}}{\nu _{l}}\leq 0, \quad
{\hbox{or}} \,\,\sum_{l=1}^{n}\frac{\lambda _{l}}{1-2\lambda _{l}}\leq 0.
\label{main5}
\end{equation}
Without loss of generality we can choose $\nu _{1}<0$ and $\nu _{l}> 0$
for all $l=2,3,\ldots ,n$. Hence, $\lambda _{1}>1/2$ and $\lambda _{l}<1/2$
for all $l=2,3,\ldots ,n$. Write:
$$
{\hbox{LHS of }}(\ref{main5}) =\frac{\lambda _{1}}{1-2\lambda _{1}}
+\sum_{l=2}^{n}\frac{\lambda _{l}}{1-2\lambda _{l}} := T_{1}+T_{2}.
\label{main7}
$$
Note that $T_{1}\leq 0$ since $\lambda _{1}>1/2$. The function
\[
f(\lambda _{i}):=\frac{\lambda _{i}}{1-2\lambda _{i}},\quad 0\leq \lambda
_{i}<1/2,
\]
is convex. Hence, $T_{2}\left( \lambda _{2},\cdots ,\lambda _{n}\right) $,
as a sum of convex functions, is convex on the simplex defined by
\[
\lambda _{2}+\cdots +\lambda _{n}\leq 1-\lambda _{1},\quad
0\leq \lambda _{i}<1/2\,,\quad \,i=2,\ldots ,n,
\]
with fixed $\lambda _{i}>1/2$.

Hence, $T_{2}$ achieves its maximum on the vertices of the simplex. One vertex
is $\left( 0,\cdots ,0\right) ,$ for which $T_{2}=0,$ and
hence $T_{1}+T_{2}<0.$
Other vertices are obtained by permutations from $\left( 1-\lambda
_{1},0,\cdots ,0\right) $ and give
$$
T_{2}=\frac{1-\lambda _{1}}{1-2(1-\lambda _{1})}=-\frac{1-\lambda _{1}}{
1-2\lambda _{1}}.
$$
Thus the maximal value of $T_1 + T_2$ is
$$
\frac{\lambda _{1}}{1-2\lambda _{1}}-\frac{
1-\lambda _{1}}{1-2\lambda _{1}}=-\frac{1-2\lambda _{1}}{1-2\lambda _{1}}=-1
$$
which proves (\ref{main5}).

To prove (\ref{main2}), using the definition (\ref{sym2a})
of elementary symmetric polynomials, we write:
$$
s_{n-k-1} (\nu_1,..,{\not{\nu_l}} \ldots, \nu_n)
= \frac{c_n}{(k-1)!}
\sum_{j_1=1\atop{j_1 \ne l}}^n  \sum_{j_2=1\atop{j_2\ne l,j_1}}^n \cdots
 \sum_{j_k=1\atop{j_k \ne l\atop{j_k\ne j_i \forall 1\le i\le k}}}^n\frac{1}{\nu_{j_1} \cdots \nu_{j_k} \nu_l}.
\nonumber\\
\label{sympol4}
$$
Here $c_n:= \nu_1,..,{{\nu_l}} \ldots, \nu_n <0$. Hence, the required
inequality \reff{main2} becomes
\be
\sum_{l=1}^n \sum_{j_1=1\atop{j_1 \ne l}}^n
\sum_{j_2=1\atop{j_2\ne l,j_1}}^n \cdots
 \sum_{j_k=1\atop{j_k \ne l\atop{j_k\ne j_i
\forall 1\le i\le k-1}}}^n
\frac{1-\nu_l}{\nu_{j_1} \cdots \nu_{j_k}\nu_l } \le 0.
\label{main73}
\ee
Once again, without loss of generality we can choose $\nu_1 <0$ and $\nu_l > 0$ for
all $l=2,3,\ldots, n$. Then
\be
{\hbox{LHS of }} \reff{main73} =\sum_{j_1=2}^n
\sum_{j_2=2\atop{j_2\ne j_1}}^n \cdots
 \sum_{j_k=2\atop{j_k\ne j_i
\forall 1\le i\le k-1}}^n
\frac{1}{\nu_{j_1} \cdots \nu_{j_k}}\left[
\frac{1 - \nu_1}{\nu_1} + \sum_{r=1}^k \frac{1 - \nu_{j_r}}{\nu_1}
+ \sum_{l=2\atop{l \ne j_i \forall 1\le i\le k}}^n
\frac{1-\nu_l}{\nu_l}\right]
\label{inf}
\ee
Equation \reff{inf} can be derived as follows:
Let 
\be
T(l,j_1,j_2,\ldots,j_k):= \frac{1 - \nu_l}{\nu_{j_1} \cdots \nu_{j_k}\nu_l }, 
\label{tee}
\ee
with $l,j_1,j_2,\ldots,j_k \in \{1,2,\ldots,n\}$ and $l,j_1,j_2,\ldots,j_k$
all different. 

Without loss of generality we can choose $\nu_1 <0$ and $\nu_l > 0$ for
all $l=2,3,\ldots, n$. Then 
\bea
{\hbox{LHS of }} \reff{main73} &=& 
\sum_{l=1}^n \sum_{j_1=1\atop{j_1 \ne l}}^n
\sum_{j_2=1\atop{j_2\ne l,j_1}}^n \cdots
 \sum_{j_k=1\atop{j_k \ne l\atop{j_k\ne j_i
\forall 1\le i\le k-1}}}^n
T(l,j_1,j_2,\ldots,j_k) \nonumber\\
&=& \sum_{j_1=2}^n
\sum_{j_2=2\atop{j_2\ne j_1}}^n \cdots
 \sum_{j_k=2\atop{j_k\ne j_i
\forall 1\le i\le k-1}}^n T(1,j_1,j_2,\ldots,j_k) + 
\sum_{l=2}^n \sum_{j_2=2\atop{j_2\ne l}}^n \cdots
 \sum_{j_k=2\atop{j_k \ne l \atop{j_k\ne j_i
\forall 1\le i\le k-1}}}^n T(l,1,j_2,\ldots,j_k) \nonumber\\
&+&  
\sum_{l=2}^n \sum_{j_1=2\atop{j_1 \ne l}}^n
\sum_{j_3=2\atop{j_3\ne j_1,l}}^n \cdots
\sum_{j_k=2\atop{j_k\ne l \atop{j_k\ne j_i
\forall 1\le i\le k-1}}}^n T(l,j_1,1,j_3, \ldots,j_k) 
+ \cdots \nonumber\\ 
&+& \sum_{l=2}^n \sum_{j_1 =2\atop{j_1 \ne l}}^n
\cdots \sum_{j_{k-1}=2\atop{j_{k-1}\ne l \atop{j_{k-1}\ne j_i
\forall 1\le i\le k-2}}}^n T(l,j_1,\ldots,j_{k-1},1)
+ \sum_{l=2}^n \sum_{j_1=2\atop{j_1\ne l}}^n \cdots
 \sum_{j_k=2 \atop{j_k \ne l \atop{j_k\ne j_i
\forall 1\le i\le k-1}}}^n  T(l,j_1,\ldots,j_k).\nonumber\\
\label{four}
\eea
Now,
\bea
& &  \sum_{l=2}^n \sum_{j_1=2\atop{j_1 \ne l}}^n \cdots
\sum_{j_{i-1}=2\atop{j_{i-1}\ne l 
\atop{j_{i-1}\ne j_r
\forall 1\le r\le i-2}}}^n
\sum_{j_{i+1}=2\atop{j_{i+1}\ne l 
\atop{j_{i+1}\ne j_r
\forall 1\le r\le i-1}}}^n\cdots 
\sum_{j_k=2\atop{j_k\ne l \atop{j_k\ne j_r
\forall 1\le r\le k-1}}}^n T(l,j_1,\ldots,j_{i-1},1,j_{i+1},\ldots,j_k) \nonumber\\
&=& \sum_{l=2}^n \sum_{j_1=2\atop{j_1 \ne l}}^n \cdots
\sum_{j_{i-1}=2\atop{j_{i-1}\ne l 
\atop{j_{i-1}\ne j_r
\forall 1\le r\le i-2}}}^n
\sum_{j_{i+1}=2\atop{j_{i+1}\ne l 
\atop{j_{i+1}\ne j_r
\forall 1\le r\le i-1}}}^n\cdots 
\sum_{j_k=2\atop{j_k\ne l \atop{j_k\ne j_r
\forall 1\le r\le k-1}}}^n 
\frac{1 - \nu_l}{\nu_1\nu_{j_1} \ldots \nu_{j_{i-1}} \nu_{j_{i+1}}\ldots \nu_{j_k}\nu_l }\nonumber\\
&=&\sum_{j_1=2}^n \cdots
\sum_{j_{i-1}=2 
\atop{j_{i-1}\ne j_r
\forall 1\le r\le i-2}}^n
\sum_{j_{i}=2 
\atop{j_{i}\ne j_r
\forall 1\le r\le i-1}}^n
\sum_{j_{i+1}=2 
\atop{j_{i+1}\ne j_r
\forall 1\le r\le i}}^n\cdots 
\sum_{j_k=2\atop{j_k\ne j_r
\forall 1\le r\le k-1}}^n 
\frac{1 - \nu_{j_i}}{\nu_1\nu_{j_1} \ldots \nu_{j_{i-1}} \nu_{j_i} \nu_{j_{i+1}}\ldots \nu_{j_k}}
\nonumber\\
&=& 
\sum_{j_1=2}^n
\sum_{j_2=2\atop{j_2\ne j_1}}^n \cdots
 \sum_{j_i=2\atop{j_i\ne j_r
\forall 1\le r\le i-1}}^n
 \sum_{j_k=2\atop{j_k\ne j_r
\forall 1\le r\le k-1}}^n
\frac{1}{\nu_{j_1} \cdots \nu_{j_k}} \left(\frac{1 - \nu_{j_i}}{\nu_1}\right).
\label{trans}
\eea
In the second last line on the RHS of \reff{trans} ,
we have changed the dummy variable from $l$ to $j_i$.
Hence, 
\bea
{\hbox{RHS of }} \reff{four} &=& \sum_{j_1=2}^n
\sum_{j_2=2\atop{j_2\ne j_1}}^n \cdots
 \sum_{j_k=2\atop{j_k\ne j_i
\forall 1\le i\le k-1}}^n
\frac{1}{\nu_{j_1} \cdots \nu_{j_k}}\left[
\frac{1 - \nu_1}{\nu_1} + \sum_{r=1}^k \frac{1 - \nu_{j_r}}{\nu_1}
+ \sum_{l=2\atop{l \ne j_i \forall 1\le i\le k}}^n
\frac{1-\nu_l}{\nu_l}\right]\nonumber\\
&=& {\hbox{RHS of }} \reff{inf}
\eea

From \reff{main5} it follows that for given $j_1, j_2, \ldots, j_k$, with
$2\le j_r \le n$ for $r=1,2, \ldots, k$, and $j_m \ne j_k$ for all $m \ne k$:
$$
\sum_{l=2\atop{l \ne j_i \forall 1\le i\le k}}^n
\frac{1-\nu_l}{\nu_l}+ \sum_{r=1}^k \frac{1 - \nu_{j_r}}{\nu_{j_r}}
+ \frac{1 - \nu_1}{\nu_1} \le 0.
\label{ineq23}
$$
Hence,
\be
\sum_{l=2\atop{l \ne j_i \forall 1\le i\le k}}^n
\frac{1-\nu_l}{\nu_l}\le - \left[\sum_{r=1}^k \frac{1 -
\nu_{j_r}}{\nu_{j_r}}
+ \frac{1 - \nu_1}{\nu_1}\right]
\label{sub22}
\ee
Substituting \reff{sub22} on the RHS of \reff{inf} yields
\bea
{\hbox{RHS of }} \reff{inf} &\le& \sum_{j_1=2}^n
\sum_{j_2=2\atop{j_2\ne j_1}}^n \cdots
 \sum_{j_k=2\atop{j_k\ne j_i
\forall 1\le i\le k}}^n
\frac{1}{\nu_{j_1} \cdots \nu_{j_k}} \left[\sum_{r=1}^k (1 - \nu_{j_r})
\bigl(\frac{1}{\nu_1} - \frac{1}{\nu_{j_r}}\bigl)\right]\nonumber\\
&=&  \sum_{j_1=2}^n
\sum_{j_2=2\atop{j_2\ne j_1}}^n \cdots
 \sum_{j_k=2\atop{j_k\ne j_i
\forall 1\le i\le k}}^n
\frac{1}{\nu_{j_1} \cdots \nu_{j_k}} \left[\sum_{r=1}^k (1 - \nu_{j_r})
\bigl(\frac{\nu_{j_r} - \nu_1}{\nu_1 \nu_{j_r}}\bigl)\right]\le 0.
\eea
This proves \reff{main73} and hence \reff{main2} for $n\ge 3$ and
all $k=1,2, \ldots, n-2$.

\section{Appendix}

A real--valued function $\Phi $ on ${\mathbf{R}}^{n}$ is said to be \emph{Schur
concave} (see \cite{bha}) if:
$$
{\underline{x}}\prec {\underline{y}}\quad \implies \Phi ({\underline{x}})\geq
\Phi ({\underline{y}}).
$$
Here the symbol ${\underline{x}}\prec {\underline{y}}$ means that ${
\underline{x}}=(x_{1},x_{2},\ldots ,x_{n})$ is \emph{majorized} by ${
\underline{y}}=(y_{1},y_{2},\ldots ,y_{n})$ in the following sense:
Let ${\underline{x}}^\downarrow$ be
the vector obtained by rearranging the coordinates of ${\underline{x}}$ in
decreasing order:
\[
{\underline{x}}^\downarrow = (x_1^\downarrow, x_2^\downarrow, \ldots,
x_n^\downarrow)\quad {\hbox{means   }} x_1^\downarrow \ge x_2^\downarrow \ge
\ldots \ge x_n^\downarrow .
\]
For ${\underline{x}}, {\underline{y}} \in {\mathbf{R}}^n$, we say that ${
\underline{x}}$ is majorized by ${\underline{y}}$ and write ${\underline{x}}
\prec {\underline{y}}$ if
$$
\sum_{j=1}^k x_{j}^\downarrow \le \sum_{j=1}^k y_{j}^\downarrow, \quad 1\le
k \le n,
$$
and
$$
\sum_{j=1}^n x_{j}^\downarrow = \sum_{j=1}^n y_{j}^\downarrow.
$$
In the simplex $\Sigma_d$, defined by the constraints \reff{nine}, 
the minimal point is $(1/d, \ldots, 1/d)$ (the baricenter
of $\Sigma_d$), and the maximal points are the permutations of $(1,0,
\ldots,0)$ (the vertices).

A differentiable function $\Phi(x_1, x_2, \ldots, x_n)$ is Schur
concave if and only if :

\begin{enumerate}
\item  {$\Phi $ is symmetric}

\item  {\
\begin{equation}
(x_{i}-x_{j})\bigl(\frac{\partial \Phi }{\partial x_{i}}-\frac{\partial \Phi
}{\partial x_{j}}\bigr)\geq 0,\quad \forall \,1\leq i,j\leq n.  \label{iden3}
\end{equation}
}
\end{enumerate}

The $l^{th}$ elementary symmetric polynomial $s_{l}$ in the variables $
x_{1},x_{2},\ldots ,x_{n}$ is defined as
\begin{eqnarray}
s_{0}(x_{1},x_{2},\ldots ,x_{n}) &=&1,  \nonumber \\
s_{l}(x_{1},x_{2},\ldots ,x_{n}) &=&\sum_{1\leq i_{1}<i_{2}\cdots <i_{l}\leq
d}x_{i_{1}}x_{i_{2}}\ldots x_{i_{l}}\quad {\hbox{for }}l=1,2,\ldots ,n.
\label{sym2a}
\end{eqnarray}
We shall use the following identities:
\begin{equation}
\frac{\partial }{\partial x_{j}}s_{k}(x_{1},x_{2},\ldots
,x_{n})=s_{k-1}(x_{1},\ldots ,{\not{x_{j}}},\ldots ,x_{n})  \label{iden1}
\end{equation}
and
\begin{equation}
s_{k}(x_{1},\ldots ,{\not{x_{i}}},\ldots ,x_{n})-s_{k}(x_{1},\ldots ,{\not
{x_{j}}},\ldots ,x_{n})=(x_{j}-x_{i})s_{k-1}(x_{1},\ldots ,{\not{x_{i}}},..,{
\not{x_{j}}},\ldots ,x_{n})  \label{iden2}
\end{equation}

\bigskip

\noindent
{\bf{Acknowledgments }} This  work was initiated when A.S.H.
was an overseas visiting scholar at
St John's College,
Cambridge.
He gratefully acknowledges the hospitality of the Statistical Laboratory
of the Centre for Mathematical Sciences. N.D. and Y.S. worked in association
with the CMI.

\end{document}